\documentstyle[12pt]{article} \voffset=0mm \headheight=0mm
\date{}
\author{Valerii Dryuma\thanks{Work supported in part by Grant RFFI, Russia-Moldova}\\[5mm]
{\it Institute of Mathematics and Informatics, AS RM,}\\[3mm] {\it
5 Academiei Street, 2028 Kishinev, Moldova},\\[3mm]{\it e-mail:
valery@dryuma.com; valdryum@gmail.com; cainar@mail.md} }
\title{ ON THE RICCI-FLAT METRIC FOR THE NAVIER-STOKES EQUATIONS}

\begin{document}
\maketitle
\date{}
\maketitle
\begin{abstract}
\ \ \   Examples of the Riccii-flat metrics associated with  the equations of Navier-Stokes are constructed.
 Their properties are investigated.
\end{abstract}

\medskip

\section{Introduction}

   Properties of solutions of the Navier-Stokes equations to the
incompressible fluid can be studied by geometric methods [1-3].

   For this purpose we rewrite the $NS$- equations
in equivalent form of conservation laws
\begin{equation}\label{law}
U_t+(U^2-\mu U_x+P)_x+(UV-\mu U_y)_y+(UW-\mu U_z)_z=0,$$$$
V_t+(UV-\mu V_x)_x+(V^2-\mu V_y+P)_y+(VW-\mu V_z)_x=0,$$$$
W_t+(UW-\mu W_x)_x+(VW-\mu W_y)_y+(W^2-\mu W_z+P)_z=0,$$$$
 U_x+V_y+W_z=0,
\end{equation}
where $U,V,W$ and $P$ are components of the velocity and the
pressure of the fluid.

   The system of equations (\ref{law}) can be considered as conditions of equality to zero the Ricci
tensor of 14- dimensional space  $D^{14}$ in local coordinates\\
$X=(x,y,z,t,\eta,\rho,m,u,v,w,p,\xi,\chi,n)=(\vec
x,t,\eta,\rho,m,\Psi_l)$ , $l=1...7$ endowed with the Riemann metric
\begin{equation}\label{metr}
 ^{14}ds^2=-2\Gamma^i_{jk}(\vec x,t)\Psi_idx^jdx^k+2 d\Psi_ldx^l.
  \end{equation}

   The metric (\ref{metr}) is the  metric of the Riemann extension [4]
of seven-dimensional space $D^7$ of affine connection in local
coordinates $(x,y,z,t,\eta,\rho,m)$ with components of connection
$\Gamma^i_{jk}(\vec x,t)$.

   In explicit form it looks as follows
\begin{equation}\label{metr1}
 ^{14} {{\it ds}}^{2}=2\,{\it dx}\,{\it du}+2\,{\it dy}\,{\it
dv}+2\,{\it dz} \,{\it dw}+\left (-V(\vec x,t)v-W(\vec
x,t)w-U(\vec x,t)u\right ){{\it dt
}}^{2}\!+\!$$$$\!+\!\left(\!-\!u\left (U(\vec x,t)\right
)^{2}\!+\!uP (\vec x,t)\!+\!u\mu\,U_x(\vec x,t)\!-\!vU(\vec x,t)V
(\vec x,t)\!-\!U(\vec x,t)p \right
)d{\eta}^{2}\!$$$$+\left(\!v\mu\,U_y(\vec x,t)\!-\! wU(\vec
x,t)W(\vec x,t)\!+\!w\mu\,U_z(\vec x,t) \right
)d{\eta}^{2}\!+\!2\,{d}^{2}\eta\,\xi\!+\!$$$$\!+\!\left (-uU(\vec
x,t)V(\vec x,t)\!+\! vP(\vec x,t)\!-\!V(\vec
x,t)p\!+\!u\mu\,V_x(\vec x,t )\!-\!wV(\vec x,t)W(\vec
x,t)\right)d{\rho}^{2}+$$$$+\left(\!v\mu\,V_y(\vec x,t
)+\!w\mu\,V_z(\vec x,t)\!-\!v\left (V(\vec x,t) \right )^{2}\right
)d{\rho}^{2}\!+\!2\,{d}^{2}\rho\,\chi\!+\!$$$$\!+\!\left (wP(\vec
x,t)\!-\!W(\vec x,t)p\!-\!w\left (W(\vec x,t)\right
)^{2}\!+\!w\mu\,W_z(\vec x,t)\!+\!v\mu\,W_y(\vec x,t)\right){{\it
dm}}^{2}\!+$$$$+\!\left(-vV(\vec x,t)W(\vec
x,t)\!+\!u\mu\,W_x(\vec x,t)\!-\!uU(\vec x,t)W(\vec x,t)\right
){{\it dm}}^{2}\!+\!2\,{\it dm}\,{\it dn}+\!2\,{\it dt}\,{\it
dp}\!.
 \end{equation}

  The main property of the space with the metric (\ref{metr1}) lies in the fact that
  it is a Ricci-flat if the functions $U,V,W$ and $P$ satisfy the NS-equations (\ref{law}).

   Despite the fact that all scalar invariants of the space $D^{14}$
    are equal to zero its geometric properties  can be studied with the help  of
    equations of geodesic and corresponding invariant differential operators.

\section{Geodesic}

      The complete system of geodesics of the metric (\ref {metr}) consists of two
parts
   $$
 \ddot x^k+\Gamma^k_{ij}\dot x^i \dot x^j=0,\quad
  \frac{\delta^2 \Psi_k}{ds^2}+R^l_{k j
i}\dot x^j \dot x^i \Psi_l=0,
 $$
  where
$$ \frac{\delta \Psi_k}{ds}=\dot \Psi_k-\Gamma^l_{jk}\Psi_l \dot x^j.
 $$

     In considered case the first group of equations is
\begin{equation} \label{geod0}
\ddot x\!+\!1/2\,U(\vec x,t)\dot t^{2}\!+\!1/2\,\dot \eta U(\vec x,t)^{2}\!-\!1/2\,\dot \eta^{2}\mu \,U_x(\vec x,t)\!-\!1/2\,\dot \eta^{2}P(\vec x,t)+$$$$+1/2\,\dot \rho^{2}U(\vec x,t)V(\vec x,t)\!-\!1/2\,\dot \rho
^{2}\mu\,V_x(\vec
x,t)\!+\!1/2\,\dot m^{2}U(\vec
x,t)W(\vec x,t)\!-$$$$-\!1/2\,\dot m(s)\mu\,W_x(\vec x,t)\!=0,
 $$
$$ \ddot y+1/2\,V(\vec x,t)\dot t^{2}+1/2\,\dot \eta(s)^{2}U(\vec x,t) V(\vec x,t)-1/2\,\dot \eta^{2}\mu\,U_y(\vec x,t)+1/2\,\dot \rho^{2}V(\vec x,t)^{2}-$$$$-1/2\,\dot \rho^{2}\mu\,V_y(\vec x,t)-1/2 \,\dot \rho(s)P(\vec x,t)+1/2\,\dot m^{2}V(\vec x,t)W(\vec x,t)-$$$$-1/2\,\dot m^{2}\mu\,W_y(\vec x,t)=0,
 $$
$$\ddot z\!+\!1/2\,W(\vec x,t)\dot t^{2}\!+\!1/2\,\dot \eta(s)^{2}U(\vec x,t) W(\vec x,t)\!-\!1/2\,\dot \eta^{2}\mu\,U_z(\vec x,t)\!+\!1/2\,\dot \rho^{2}V(\vec x,t)W(\vec x,t)\!-\!$$$$\!-\!1/2\,\dot \rho^{2}\mu\,V_z(\vec x,t)\!+\!1/2\,\dot m^{2}W(\vec x,t)^{2}\!-\!1/2\,\dot m^{2}\mu\,W_z(\vec x,t)\!-\!1/2\,\dot m^{2}P(\vec x,t) =0,
$$
 $$ \ddot t+1/2\,U(x,y,z,t)\dot \eta^{2}+1/2\,V(x,y,z,t)\dot \rho^{2}+1/2\,W(x,y,z,t)\dot m^{2}=0,
 $$
   $$
 \ddot \eta=0,~~ \ddot \rho(s)=0,~~
\ddot m(s)=0.
\end{equation}

    In  particular case of $2D$-potential flow $$U=\phi_y,~ V=-\phi_x,~ W=0, ~P=Q(x,y,t)$$ the system (\ref{geod0})
    takes the form
$$
2\,\ddot x+\phi_y \dot t^{2}+{{\it
\alpha1}}^{2}\phi_y^
{2}-{{\it \alpha_1}}^{2}\mu\,\phi_{xy}-{{\it \alpha_1}}^{2}Q-{{\it \alpha_2}}^{2}\phi_y\phi_x+{{\it \alpha_2}}^{2}\mu\,\phi_{xx}=0,
$$
$$2\,\ddot y-\phi_x \dot t^{2}-{{\it
\alpha_1}}^{2}\phi_y\phi_x-{{\it \alpha_1}}^{2}\mu\,\phi_{yy}+{{\it \alpha_2}}^{2}
\phi_x^{2}+{{\it
\alpha_2}}^{2}\mu\,\phi_{xy}-{{\it \alpha_2}}^{2}Q=0,$$$$
2\,\ddot z-Q{{\it \alpha_3}}^{2}=0,~~
2\,\ddot t+\phi_y{{\it \alpha_1}}^{2}-\phi_x{{\it \alpha_2}}^{2}=0,
$$
$$
\eta(s)=\alpha_1s,~\rho(s)=\alpha_2s,~ m(s)=\alpha_3s.
$$
{\bf Remark.}
    The coefficients of the system (\ref{geod0}) are the
 components $\Gamma^k_{ij}$ of affine connection of the
seven-dimensional manifold in the local coordinates $(\vec
x,t,\eta,\rho,m)$.

    It is a Ricci-flat $$ R_{ij}=\partial_k
\Gamma^k_{ij}-\partial_i
\Gamma^k_{kj}+\Gamma^k_{kl}\Gamma^l_{ij}-\Gamma^k_{im}\Gamma^m_{kj}=0
$$ on solutions of the NS-equations (\ref{law})  and its properties can be studied independently of the
enclosing $14$-dimensional Riemann space with the metric
(\ref{metr1}).

    Linear part of geodesic has the form of linear system of
    equations  with variable coefficients
    $$
    \ddot \Psi_i=A^k_i \dot \Psi_k+B^k_i \Psi_k,
 $$
 where $\Psi_k=[u,v,w,p,\xi,\chi,n]$ is vector-functions and $ A^k_i= A^k_i(\vec x,t)$ and $ B^k_i=B^k_i(\vec x,t)$  are the matrix-functions depending on coordinates $X^a=(\vec x,t)$ .

   Properties of the seven-dimension of the space of affine connection can be studied with the help of solutions of a system of linear partial differential equations with respect to
the components of the vector of motions $\omega^k (\vec x, t,\eta,\rho, m)$ of the space
$$
\partial^2_{bc}\omega^a+\omega^k\partial_k \Gamma^a_{bc}+\partial_b\omega^k \Gamma^a_{kc}+\partial_c\omega^k\Gamma^a_{bk}-\partial_k\omega^a\Gamma^k_{bc}=0.
$$

\section{Parameters of Beltrami}

  The functions of coordinates $\psi(x^k)$ which defined by the formulas
\begin{equation} \label{Lap-Bel2}
\Delta_2\psi=g^{ij}\left(\frac{\partial^2 \psi}{\partial x^i
\partial x^j}-\Gamma^k_{ij}\frac{\partial\psi}{\partial
x^k}\right),
\end{equation}
and
\begin{equation} \label{Lap-Bel1}
\Delta_1\psi=g^{ij}\frac{\partial \psi}{\partial x^i}
\frac{\partial \psi}{\partial x^j}
\end{equation}
are the invariants of the space.

   Solutions of the equations $$\Delta_2\phi=0,~ \Delta_1 \phi=0$$ can be used to
     the study properties of solutions of the $NS$ -equations.

     As an example, consider the two-dimensional potential flow of fluid.

     The  metric of associated Riemann space in this case has the form
\begin{equation} \label{poten}
   {{\it ds}}^{2}=2\,{\it dx}\,{\it du}+2\,{\it dy}\,{\it dv}+2\,{\it dz}
\,{\it dw}+\left (-\phi_y
u+\phi_x v
\right ){{\it dt}}^{2}+2\,{\it dt}\,{\it dp}+$$$$+\left (-u\phi_y^{2}+u\mu\,\phi_{xy}+uQ(x,y,t)+v\phi_y \phi_x+v\mu\,\phi_{yy}-\phi_y p\right )
{{\it d\eta}}^{2}+$$$$+2\,{\it d\eta}\,{\it d\xi}+\left (u\phi_y \phi_x-u\mu\,\phi_{xx}-v\phi_x^{2}-v
\mu\,\phi_{xy}+vQ(x,y,t
)+\phi_x p\right ){{
\it d\rho}}^{2}+$$$$+2\,{\it d\rho}\,{\it d\chi}+wQ(x,y,t){{\it dm}}^{2}+2\,{
\it dm}\,{\it dn}.
\end{equation}

    Components of the Ricci-tensor of the metric (\ref{poten})
are equal to zero $R_{\eta \eta}=0$,  $R_{\rho \rho}=0$
on solutions of $2DNS$-equations
$$
\phi_y \phi_{xy}-\mu\,\phi_{xxy}-Q_x-\phi_{yy}\phi_x-\mu\,\phi_{yyy}+
\phi_{yt}=0,
$$
$$
-\phi_y \phi_{xx}+\mu\,\phi_{xxx}+\mu\,\phi_{xyy}-Q_y-\phi_{xt}+\phi_{xy}\phi_x=0,
$$
where the function $\phi(x,y,t)$ satisfies the condition of compatibility
\begin{equation} \label{flow}
(\phi_{xx}+\phi_{yy})_t+\phi_y(\phi_{xx}+\phi_{yy})_x-\phi_x(\phi_{xx}+\phi_{yy})_y-\mu \Delta(\phi_{xx}+\phi_{yy})=0.
\end{equation}

    Here is an example of the solution of equation
 \begin{equation} \label{Lap-Bel3}
g^{ij}\frac{\partial \psi}{\partial x^i}
\frac{\partial \psi}{\partial x^j}-1=0.
\end{equation}

   As is known [5] with the help of solutions equation (\ref{Lap-Bel3}) can be studied
     geodesics of the metric for  arbitrary Riemann space.

     For the metric (\ref{poten}) the equation (\ref{Lap-Bel3}) takes the form
 \begin{equation} \label{Roma1}
2\,\psi_x\psi_u+2\,\psi_y\psi_v+2
\,\psi_z\psi_w+2\,\psi_t\psi_p+2\,\psi_\eta\psi_\xi+2\,\psi_\rho\psi_\chi+
2\,\psi_m\psi_n+\psi_p^{2}\phi_yu-$$$$-\psi_p^{2}\phi_x v+\psi_\xi^{2}u\phi_y
^{2}-\psi_{\xi}^{2}u\mu\,\phi_{xy}-\psi_{\xi}^{2}uQ-\psi_\xi^{2}v\phi_y\phi_x-\psi_\xi^{2}v\mu\,\phi_{yy}+\psi_\xi^{2}\phi_yp-$$
$$-\psi_\chi^{2}u\phi_y\phi_x+\psi_\chi^{2}u\mu\,\phi_{xx}+\psi_\chi^{2}v\phi_x^{2}+\psi_\chi^{2}v\mu\,\phi_{xy}-\psi_\chi^{2}vQ-\psi_\chi^{2
}\phi_xp-wQ\psi_n^{2}-1=0.
\end{equation}

     After separation of variables in the equation (\ref{Roma1}) we find
     $$
     \psi(x,y,z,t,\eta,\rho,m,u,v,w,p,\xi,\chi,n)={\it c}_{{3}}z+{\it c
}_{{5}}\eta+{\it c}_{{6}}\rho+{\it c}_{{7}}m+{\it c}_{{12}}\xi+$$$$+{
\it c}_{{13}}\chi+{\it c}_{{14}}n+F(x,y,t,u,v,w,p)
$$
where the function $F(x,y,t,u,v,w,p)$ satisfies the equation
\begin{equation} \label{Roma2}
2\,F_x F_u+2\,F_yF_v+2\,{
\it c}_{{3}}F_w+2\,F_tF_p+2\,{
\it c}_{{5}}{\it c}_{{12}}+2\,{\it c}_{{6}}{\it c}_{{13}}+2\,{
\it c}_{{7}}{\it c}_{{14}}+$$$$+F_p^{2}\phi_yu-F_p^{2}\phi_xv+{{\it c}_{{12}}}^{2}u\phi_y^{2}-{{\it c}_{{12}}}^{2}
u\mu\,\phi_{xy}-{{\it
c}_{{12}}}^{2}uQ-{{\it c}_{{12}}}^{2}v\phi_y\phi_x-$$$$-{{\it c}_{{12}}}^{2}v\mu\,\phi_{yy}+{{\it c}_{{12}}}^{2}\phi_yp-{{\it c}_{{13}}}^{2}u
\phi_y\phi_x+{{\it c}_{{13}}}^{2}u\mu\,\phi_{xx}+{{\it c}_{{13}}}^{2}v
\phi_x^{2}+$$$$+{{\it
c}_{{13}}}^{2}v\mu\,
\phi_{xy}-{{\it c}_{{13}}}^{2}vQ-{{\it c}_{{13}}}^{2}
\phi_xp-wQ{{
\it c}_{{14}}}^{2}-1=0.
\end{equation}

    Using presentation the function $F(x,y,t,u,v,w,p)$ in the form
    $$
    F(x,y,t,u,v,w,p)=A(x,y,t)p+uB(x,y,t),
{\it c}_{{14}}=0,
{\it c}_{{13}}=0,
{\it c}_{{12}}=1/2\,{{\it c}_{{5}}}^{-1}
$$
we get over determinant system of equations to the functions $A(x,y,t)$, $B(x,y,t)$
\begin{equation} \label{Roma3}
8\,B{{\it c}_{{5}}}^{2}B_x
+8\,A{{\it c}_{{5}}}^{2}B_t+4\,A^{2}\phi_y{{\it c}_{{5}}}^{2}+\phi_y^{2}-\mu\,\phi_{xy}-Q=0,
$$
$$
8\,B{{\it c}_{{5}}}^{2}A_x+8\,A{{\it c}_{{5}}}^{2}A_t+\phi_y=0,~~
-4\,A^{2}
\phi_x{{\it c}_{{5}}}^{2}-\phi_y\phi_x-\mu\,\phi_{yy}=0.
\end{equation}

     From conditions of compatibility we find that the system (\ref{Roma3}) has solutions
     if the functions $\phi(x,y,t)$, $Q(x,y,t)$ satisfy to the relation
\begin{equation} \label{euler}
H(\phi,\phi_x,\phi_y,\phi_t,...Q)=0
\end{equation}
containing $275$ summands.

    In particular, for the Euler system of equations, ($\mu=0$)  the relation (\ref{euler}) takes the form
$$
\phi_{yt}\phi_{xy}^{2}\phi_y-4\,\phi_{yt}\phi_y\phi_{xy}
\phi_{xyt}+2\,\phi_{y}\phi_{xxy}\phi_{yt}^{2}-4
\,\phi_{yt}\phi_y^{2}\phi_{xxy}-$$$$-3\,
\phi_{xy}^{2}\phi_y^{2}+4\,
\phi_y^{2}\phi_{xy}\phi_{xyt}+2\,\phi_y^{3}\phi_{xxy}+2\,\phi_y\phi_{xy}^{2}\phi_{ytt}-Q\phi_{xy}^{3}
=0.
$$

\end{document}